\documentclass{jfm}

\usepackage{graphicx}
\usepackage{epstopdf}
\usepackage{amsmath}
\usepackage{amsfonts}
\usepackage{amssymb}
\usepackage{mathrsfs}
\usepackage{subfigure}
\usepackage{natbib}
\usepackage{bm}
\usepackage{color}
\usepackage{units}
\usepackage[dvipsnames]{xcolor}
\usepackage{multirow}
\usepackage{romannum}

\ifCUPmtlplainloaded \else
  \checkfont{eurm10}
  \iffontfound
    \IfFileExists{upmath.sty}
      {\typeout{^^JFound AMS Euler Roman fonts on the system,
                   using the 'upmath' package.^^J}%
       \usepackage{upmath}}
      {\typeout{^^JFound AMS Euler Roman fonts on the system, but you
                   dont seem to have the}%
       \typeout{'upmath' package installed. JFM.cls can take advantage
                 of these fonts,^^Jif you use 'upmath' package.^^J}%
      }
  \else
  \fi
\fi

\ifCUPmtlplainloaded \else
  \checkfont{msam10}
  \iffontfound
    \IfFileExists{amssymb.sty}
      {\typeout{^^JFound AMS Symbol fonts on the system, using the
                'amssymb' package.^^J}%
       \usepackage{amssymb}%

      }{}
  \fi
\fi

\ifCUPmtlplainloaded \else
  \IfFileExists{amsbsy.sty}
    {\typeout{^^JFound the 'amsbsy' package on the system, using it.^^J}%
     \usepackage{amsbsy}}
    {\providecommand\boldsymbol[1]{\mbox{\boldmath $##1$}}}
\fi

\newcommand{\aavg}[1]{\left\langle #1 \right\rangle}

\newcommand{\ParDeri}[2]{\frac{\partial #1}{\partial #2}}
\newcommand{\ParDeriH}[2]{(\partial #1/\partial #2)}

\def\Ecal{\mathcal{E}}
\def\Scal{\mathcal{S}}
\def\Kcal{\mathcal{K}}

\def \d{\textrm{d}}

\usepackage{psfrag,graphicx,epsfig,color}

\newlength{\SGfigurewidth}
\newlength{\SGlabelwidth}

\newcommand{\ie}{\textit{i.e.}}
\newcommand{\eg}{\textit{e.g.}}

\newcommand{\RNum}[1]{\uppercase\expandafter{\romannumeral #1\relax}}

\title[Dilatational contribution to energy flux in compressible turbulence]
{The contribution of dilatational motion to energy flux in homogeneous compressible turbulence}

\author[C. Luo, L. Fang, J. Fang, A. Pumir, H. Xu and P.-F. Yang]
{Chensheng Luo$^{1,2,3}$, Le Fang$^{3,2}$, Jian Fang$^{4}$, Haitao Xu$^{5}$, Alain Pumir$^{6}$ and Ping-Fan Yang$^{1,5,7}$\thanks{Email address for correspondence: yangpingfan@nwpu.edu.cn}}

\affiliation{
$^1$ School of Aeronautics and Institute of Extreme Mechanics, Northwestern Polytechnical University, Xi'an, 710072, PR China
\\[\affilskip]
$^2$ Research Institute of Aero-Engine, Beihang University, Beijing 100191, PR China
\\[\affilskip]
$^3$ Laboratory of Complex System, Ecole Centrale de P\'ekin/School of General Engineering, Beihang University, Beijing 100191, PR China
\\[\affilskip]
$^4$ Scientific Computing Department, STFC Daresbury Laboratory, Warrington WA4 4AD, UK
\\[\affilskip]
$^5$  Center for Combustion Energy and School of Aerospace Engineering\\Tsinghua University, Beijing, 100084, PR China
\\[\affilskip]
$^6$ Laboratoire de Physique, Ecole Normale Sup\'erieure de Lyon, CNRS\\Universit\'e de Lyon, Lyon,  F-69007 France
\\[\affilskip]
$^7$ National Key Laboratory of Aircraft Configuration Design, Xi’an 710072, China
}

\pubyear{}
\volume{}
\pagerange{}
\date{9 October 2024}
\begin{document}
\maketitle

\begin{abstract}
We analyze the energy flux in compressible turbulence by generalizing the exact decomposition recently proposed by Johnson (Phys. Rev. Lett., vol. 124, 2020. 104501) to study incompressible turbulent flows. This allows us to characterize the effect of dilatational motion on the inter-scale energy transfer in three-dimensional compressible turbulence. Our analysis reveals that the contribution of dilatational motion to energy transfer is due to three different physical mechanisms: the interaction between dilatation and strain, between dilatation and vorticity, and the self-interaction of dilatational motion across scales. By analyzing numerical simulations of flows at moderate turbulent Mach numbers ($Ma_t \lesssim 0.3$), we validate our theoretical derivations and provide a quantitative description of the role of dilatational motion in energy transfer. In particular, we determine the scaling dependence of the dilatational contributions on the turbulent Mach number. Moreover, our findings reveal that the eddy-viscosity assumption often used in large-eddy simulations, in the spirit of the approach used for incompressible flows, effectively neglects the interaction between solenoidal-dilatational energy transfer and overestimate dilatational effects.
\end{abstract}

\maketitle

\section{Introduction}
\label{sec:intro}

Fluid turbulence is characterized by a wide range of spatial scales, and understanding how energy is distributed and transferred across scales is a fundamental problem. In three-dimensional (3D) incompressible turbulence, energy undergoes a cascade from the largest scales of the system, where energy is injected, to the smallest scales, where it is dissipated~\citep{Richardson1922, MoninYaglom1975, TennekesLumley1975, Frisch95, Pope2000, Davidson2015}. One of the clearest signatures of the cascade comes from the third-order velocity structure function, which satisfies the K\'arm\'an-Howarth equation~\citep{KarmanHowarth1938HIT}. This leads to the celebrated four-fifths law for the third-order structure function~\citep{K41a,K41b}, which indicates energy transfer from large to small scale in the inertial range of turbulence. An alternative approach to quantify energy transfer consists in studying the problem in Fourier space, which leads to the K\'arm\'an-Lin equation for the energy spectrum~\citep{Lin1947,VonkarmanLin1951}, or the subgrid-scale (SGS) stress~\citep{Eyink2006, Ouellette2018TensorGeometry, Ouellette2020GeometricConstraints, Dong2020TransferShearTur} in the filtered energy equation within the Large Eddy Simulation (LES) formalism~\citep{Germano1992filtering, Meneveau_Katz_ARFM_2000, Moser2021ARFM,SagautLES2006}. Recent attempts have been made to relate the descriptions in Fourier and in real space~\citep{Goto:2017,McKeown:2023}.

Energy transfer is of broader interest than just for incompressible turbulence. Here, we focus on compressible flows, which are of interest not only for fundamental reasons but also for applications to supersonic flows~\citep{kovasznay1953turbulence,spina1994physics,modesti2022direct}, ramjets~\citep{waltrup1987liquid,billig1988combustion,curran1996fluid}, volcanic eruptions~\citep{ogden2008effects} and interstellar turbulences~\citep{elmegreen2004interstellar,scalo2004interstellar,chernyshov2010weakly,ferrand2020compressible}. Numerous studies have investigated the statistics and scaling in 
compressible turbulence~\citep{Samtaney2001, Pirozzoli2004DecayingDNS, Wang2012JFM, Wang2012PRL, Wang2017PRF, JagannathanDonzis2016scaling,Donzis2020scaling}, especially the inter-scale energy transfer~\citep{Bataille1999energytransfer, Aluie2011PRL, Hussein2012ApJL, Aluie2013scaledecom, Eyink2018Cascade, Schmidt2019energytransfer}. Conceptually, one may use approaches in terms of structure function~\citep{Falkovich2010correlation, Galtier2011PRL, Falkovich2012correlation, Kritsuk2013cascade, Galtier2014polytropic, Lai2018KHMeq, Lindborg2019, Hellinger2021CompressibleMHD, Hellinger2021PRF}, or focus on the spectral approach~\citep{Kida1992DecayingDNS, MiuraKida1995exchange}. 

In general, the introduction of compressibility adds considerable complexity compared to the incompressible case, and it is challenging to extend known theoretical results to compressible turbulence. Difficulties arise as a consequence of varying density, and generally because the velocity gradient tensor is not divergence-free. Generalizations of the K\'arm\'an-Howarth equation to compressible flows require density-weighted velocity and velocity structure function~\citep{Kida1992DecayingDNS, Falkovich2010correlation, Galtier2011PRL, Falkovich2012correlation, Lai2018KHMeq, Hellinger2021PRF}, which considerably complicates the mathematical expressions and blurs the physical interpretations of the various terms in the generalized K\'arm\'an-Howarth equation. Additionally, the presence of non-zero divergence in the velocity gradient prevents us from obtaining a straightforward relation akin to the four-fifths law observed in incompressible turbulence. These challenges hinder our understanding of how density fluctuations and dilatational motion impact energy transfer in compressible turbulence. Similar difficulties arise when attempting a description in Fourier space. To fully utilize the theoretical findings from incompressible turbulence, a common approach is to decompose the compressible velocity field into solenoidal (divergence-free) and dilatational (compressive) components using Helmholtz decomposition. However, the fluctuation of density complicates this decomposition. Some studies~\citep{Fauchetphdthesis, Bataille1999energytransfer,Falkovich2017loop,SagautCambon2018} apply the Helmholtz decomposition directly to the velocity field $\boldsymbol{u}$, resulting in a solenoidal mode with zero divergence, but excluding density variations from the kinetic energy spectrum. Other studies~\citep{Kida1992DecayingDNS, MiuraKida1995exchange,Wang2013Cascade} decompose the density-weighted velocity field $\boldsymbol{w}=\sqrt{\rho}\boldsymbol{u}$, which takes into account density fluctuations and ensures positive-definiteness, but yields a solenoidal velocity field that is not strictly divergence-free, complicating the physical interpretation of the density-weighted velocity $\boldsymbol{w}$.

In this work, we address these technical challenges by applying a filtering approach and extending the formalism recently proposed by~\cite{Johnson2020PRL,Johnson2021JFM} to compressible turbulence, and unveil the physical mechanism associated with dilatational motion. \cite{Johnson2020PRL} previously showed that the SGS stress can be expressed as an integral of the filtered velocity gradient across all sub-filter scales. This approach identifies three mechanisms contributing to energy flux: strain-self amplification, vortex stretching, and mixed strain-vorticity interaction. By conducting Direct Numerical Simulations (DNSs), ~\cite{Johnson2020PRL} revealed that strain-self amplification dominates the energy cascade process. This formalism has also been extended to sheared, stably stratified turbulence~\citep{Bragg2022SSAT} and helical turbulence~\citep{JohnsonLinkmann2023helicity}. Here, we show that when applied to compressible turbulence, this approach helps identify how dilatational motion contributes to the energy flux. 

In \S \ref{sec:theory}, we will first generalize the relation between SGS stress and filtered velocity gradient for incompressible turbulence with a Gaussian filter, introduced by~\cite{Johnson2020PRL,Johnson2021JFM}, to the case of compressible turbulence using the Favre filtering, and introduce new mechanisms associated with dilatational motion. Then, in \S \ref{sec:DNS} and \S \ref{sec:num_energy_flux}, we will use DNS data to verify our theoretical derivation and analyze the dilatational contribution to energy transfer. Finally, in \S \ref{sec:conclution}, we summarize the results of this work.

\section{Relation between SGS stress and filtered velocity gradient in compressible turbulence}
\label{sec:theory}

We first start from the subgird-scale (SGS) stress term in compressible turbulence:
\begin{equation}
\tau_{ij}^\ell = \overline{\rho}^\ell\widetilde{u_i u_j}^\ell - \overline{\rho}^l\widetilde{u}_i^\ell\widetilde{u}_j^\ell = \overline{\rho u_i u_j}^\ell - \frac{1}{\overline{\rho}^\ell} \overline{\rho u_i}^\ell \overline{\rho u_j}^\ell,
\label{eq:def_SGS}
\end{equation}
where the symbol $\overline{\cdot}^\ell$ and $\widetilde{\phi}^\ell = \overline{\rho \phi}^\ell/\overline{\rho}^\ell$ denote the spatial and Favre filtering at scale $\ell$, see, \eg, \cite{Chai2012LES} and \cite{QiLiHuYu2022LES}. If we choose the spatial filter to be Gaussian, then we have the following identity \citep{Johnson2020PRL,Johnson2021JFM}:
\begin{equation}
\frac{\partial \overline{\phi}^\ell}{\partial \ell^2} = \frac{1}{2}\nabla^2 \overline{\phi}^\ell.
\label{eq:GaussDiffusion}
\end{equation}
To address the effect of density fluctuation, we need to evaluate derivatives related to Favre filtering.  We first look at $\frac{\partial \widetilde{\phi}^\ell}{\partial \ell^2}$:
\begin{align}
  \frac{\partial \widetilde{\phi}^\ell}{\partial \ell^2} &= \frac{\partial }{\partial \ell^2} \left(\frac{\overline{\rho \phi}^\ell}{\overline{\rho}^\ell}\right) = \frac{1}{2}\left[ \frac{1}{\overline{\rho}^\ell} \nabla^2 \left(\overline{\rho \phi}^\ell\right) - \frac{\overline{\rho \phi}^\ell}{(\overline{\rho}^\ell)^2}  \nabla^2 \left(\overline{\rho}^\ell\right)   \right].
  \label{eq:DtildephiDl2}
  \end{align}
Next we consider $\nabla^2 \widetilde{\phi}^\ell$:
\begin{align}
  \nabla^2 \widetilde{\phi}^\ell = \nabla^2 \left( \frac{\overline{\rho \phi}^\ell}{\overline{\rho}^\ell} \right) = \frac{1}{\overline{\rho}^\ell} \nabla^2 \left(\overline{\rho \phi}^\ell\right) - \frac{2}{(\overline{\rho}^\ell)^2} \frac{\partial \overline{\rho}^\ell}{\partial x_k} \frac{\partial \overline{\rho \phi}^\ell}{\partial x_k} + \frac{2\overline{\rho \phi}^\ell}{(\overline{\rho}^\ell)^3} \frac{\partial \overline{\rho}^\ell}{\partial x_k} \frac{\partial \overline{\rho }^\ell}{\partial x_k}  -\frac{\overline{\rho \phi}^\ell}{(\overline{\rho}^\ell)^2}  \nabla^2 \left(\overline{\rho}^\ell\right).
  \label{eq:Nablatildephi}
  \end{align}
Subtracting one-half of Eq.~\eqref{eq:Nablatildephi} from Eq.~\eqref{eq:DtildephiDl2} yields:
\begin{align}
  \frac{\partial \widetilde{\phi}^\ell}{\partial \ell^2} - \frac{1}{2} \nabla^2 \widetilde{\phi}^\ell = \frac{1}{(\overline{\rho}^\ell)^2} \frac{\partial \overline{\rho}^\ell}{\partial x_k} \frac{\partial \overline{\rho \phi}^\ell}{\partial x_k} - \frac{\overline{\rho \phi}^\ell}{(\overline{\rho}^\ell)^3} \frac{\partial \overline{\rho}^\ell}{\partial x_k} \frac{\partial \overline{\rho }^\ell}{\partial x_k}  
  = \frac{1}{\overline{\rho}^\ell} \frac{\partial \overline{\rho}^\ell}{\partial x_k} \frac{\partial \widetilde{\phi}^\ell}{\partial x_k}.
  \label{eq:tildephiEq}
  \end{align}
Now with the help of Eq.~\eqref{eq:tildephiEq}, we evaluate $\frac{\partial \tau_{ij}^\ell}{\partial \ell^2}$:
\begin{align}
\frac{\partial \tau_{ij}^\ell}{\partial \ell^2} &= \frac{\partial \overline{\rho}^\ell}{\partial \ell^2} \widetilde{u_i u_j}^\ell + \overline{\rho}^\ell \frac{\partial \widetilde{u_i u_j}^\ell}{\partial \ell^2} - \frac{\partial \overline{\rho}^l}{\partial \ell^2} \widetilde{u}_i^\ell \widetilde{u}_j^\ell - \overline{\rho}^l \frac{\partial \widetilde{u}_i^\ell}{\partial \ell^2} \widetilde{u}_j^\ell - \overline{\rho}^l \widetilde{u}_i^\ell \frac{\partial \widetilde{u}_j^\ell}{\partial \ell^2}   \notag \\
&= \left[\frac{1}{2} \nabla^2 \left(\overline{\rho}^\ell\right) \widetilde{u_i u_j}^\ell\right] + \left[\frac{1}{2}  \overline{\rho}^\ell \nabla^2 \widetilde{u_i u_j}^\ell + \frac{\partial \overline{\rho}^\ell}{\partial x_k}\frac{\partial \widetilde{u_i u_j}^\ell}{\partial x_k}\right] - \left[\frac{1}{2} \nabla^2 \left(\overline{\rho}^\ell \right) \widetilde{u_i}^\ell \widetilde{u_j}^\ell\right] \notag \\ & - \left[\frac{1}{2}\overline{\rho}^\ell \nabla^2 \left(\widetilde{u_i}^\ell\right) \widetilde{u_j}^\ell + \frac{\partial \overline{\rho}^\ell}{\partial x_k}\frac{\partial \widetilde{u_i}^\ell}{\partial x_k} \widetilde{u_j}^\ell\right] - \left[\frac{1}{2}\overline{\rho}^\ell \nabla^2 \left(\widetilde{u_j}^\ell\right) \widetilde{u_i}^\ell + \frac{\partial \overline{\rho}^\ell}{\partial x_k}\frac{\partial \widetilde{u_j}^\ell}{\partial x_k} \widetilde{u_i}^\ell\right].
\label{eq:DtauijDl2A}
\end{align}
On the other hand, we also have:
\begin{align}
 \frac{1}{2} \nabla^2 \tau_{ij}^\ell &= \frac{1}{2}\frac{\partial}{\partial x_k} \left(\frac{\partial \overline{\rho}^\ell}{\partial x_k} \widetilde{u_i u_j}^\ell + \overline{\rho}^\ell \frac{\partial \widetilde{u_i u_j}^\ell}{\partial x_k} - \frac{\partial \overline{\rho}^l}{\partial x_k} \widetilde{u}_i^\ell \widetilde{u}_j^\ell - \overline{\rho}^l \frac{\partial \widetilde{u}_i^\ell}{\partial x_k} \widetilde{u}_j^\ell - \overline{\rho}^l \widetilde{u}_i^\ell \frac{\partial \widetilde{u}_j^\ell}{\partial x_k}   \right)  \notag \\
 &= \frac{1}{2} \nabla^2 \left(\overline{\rho}^\ell\right) \widetilde{u_i u_j}^\ell + \frac{1}{2}  \overline{\rho}^\ell \nabla^2 \widetilde{u_i u_j}^\ell + \frac{\partial \overline{\rho}^\ell}{\partial x_k}\frac{\partial \widetilde{u_i u_j}^\ell}{\partial x_k} - \frac{1}{2} \nabla^2 \left(\overline{\rho}^\ell \right) \widetilde{u_i}^\ell \widetilde{u_j}^\ell  - \frac{\partial \overline{\rho}^\ell}{\partial x_k}\frac{\partial \widetilde{u_i}^\ell}{\partial x_k} \widetilde{u_j}^\ell \notag \\ & - \frac{\partial \overline{\rho}^\ell}{\partial x_k}\frac{\partial \widetilde{u_j}^\ell}{\partial x_k} \widetilde{u_i}^\ell - \overline{\rho}^\ell \frac{\partial \widetilde{u_i}^\ell}{\partial x_k}\frac{\partial \widetilde{u_j}^\ell}{\partial x_k}  - \frac{1}{2}\overline{\rho}^\ell \nabla^2 \left(\widetilde{u_i}^\ell\right) \widetilde{u_j}^\ell - \frac{1}{2}\overline{\rho}^\ell \nabla^2 \left(\widetilde{u_j}^\ell\right) \widetilde{u_i}^\ell.
\label{eq:nabla2tauij}
\end{align}
Subtracting Eq.~\eqref{eq:nabla2tauij} from Eq. \eqref {eq:DtauijDl2A} yields:
\begin{equation}
\frac{\partial \tau_{ij}^\ell}{\partial \ell^2} = \frac{1}{2} \nabla^2 \tau_{ij}^\ell + \overline{\rho}^\ell \frac{\partial \widetilde{u_i}^\ell}{\partial x_k}\frac{\partial \widetilde{u_j}^\ell}{\partial x_k}.
\label{eq:tauij_Eq}
\end{equation}
The solution of Eq.~\eqref{eq:tauij_Eq} is:
\begin{align}
\tau_{ij}^\ell &= \int_0^{\ell^2} \overline{\overline{\rho}^{\sqrt \alpha} \frac{\partial \widetilde{u_i}^{\sqrt \alpha}}{\partial x_k}\frac{\partial \widetilde{u_j}^{\sqrt \alpha}}{\partial x_k}}^{\sqrt {\ell^2 - \alpha}} \d\alpha, \notag \\
&= \int_0^{\ell^2} \overline{\overline{\rho}^{\sqrt \alpha}  \widetilde{A}_{ik}^{\sqrt \alpha} \widetilde{A}_{jk}^{\sqrt \alpha} }^{\sqrt {\ell^2 - \alpha}}\d\alpha,
\label{eq:tauij_solution}
\end{align}
where $\widetilde{A}_{ij}^{\ell} \equiv \partial \widetilde{u_i}^{\ell}/\partial x_j $ is the gradient of the Favre-filtered velocity, referred to in the following as filtered velocity gradient. Its diverge-free, symmetric and anti-symmetric parts are defined by: $\widetilde{S}_{ij}^{\ell} = \left( \widetilde{A}_{ij}^{\ell} + \widetilde{A}_{ji}^{\ell}\right)/2 -\widetilde{A}_{ll}^{\ell} \delta_{ij}/3  $ and $\widetilde{W}_{ij}^{\ell} = \left( \widetilde{A}_{ij}^{\ell} - \widetilde{A}_{ji}^{\ell}\right)/2$, respectively. Note that Eq.~\eqref{eq:tauij_solution} in our text and Eq.~(10) in~\cite{Johnson2020PRL} share a similar form, the differences being that a factor of density appears, and the substitution of standard filtering with Favre filtering for the filtered velocity gradient in the compressible case considered here. 

The quantity $\Pi^\ell \equiv \tau_{ij}^\ell \frac{\partial \widetilde{u_i}^{\ell}}{\partial x_j}$ represents the energy transfer across length scale $\ell$. By substituting the velocity gradient decomposition $\widetilde{A}_{ij} = \widetilde{S}_{ij}+\widetilde{W}_{ij}+\widetilde{A}_{ll}\delta_{ij}/3$ into Eq.~\eqref{eq:tauij_solution}, the inter-scale energy transfer term $\Pi^{\ell}$ decomposes as follows:
\begin{align}
  \Pi^{\ell} =\underbrace{\Pi^{\ell}_{s1} + \Pi^{\ell}_{s2} + \Pi^{\ell}_{s3}}_{\Pi^{\ell}_s} + \underbrace{\Pi^{\ell}_{m1} + \Pi^{\ell}_{m2} + \Pi^{\ell}_{m3}}_{\Pi^{\ell}_m} + \Pi^{\ell}_d,
\label{eq:Pi_solution_all} 
\end{align}
where,
\begin{subequations}
\begingroup
\allowdisplaybreaks
\begin{gather}
  \Pi^{\ell}_{s1} = \int_0^{\ell^2} \d\alpha \overline{\overline{\rho}^{\sqrt \alpha}  \widetilde{S}_{ik}^{\sqrt \alpha} \widetilde{S}_{jk}^{\sqrt \alpha} }^{\sqrt {\ell^2 - \alpha}} \widetilde{S}_{ij}^{\ell}, \label{eq:Pi_s1} \\
  \Pi^{\ell}_{s2} = - \int_0^{\ell^2} \d\alpha \overline{\overline{\rho}^{\sqrt \alpha}  \widetilde{W}_{ik}^{\sqrt \alpha} \widetilde{W}_{kj}^{\sqrt \alpha} }^{\sqrt {\ell^2 - \alpha}} \widetilde{S}_{ij}^{\ell}, \label{eq:Pi_s2} \\
  \Pi^{\ell}_{s3} = \int_0^{\ell^2} \d\alpha \overline{\overline{\rho}^{\sqrt \alpha} \left(\widetilde{S}_{kj}^{\sqrt \alpha} \widetilde{W}_{ik}^{\sqrt \alpha}  - \widetilde{S}_{ik}^{\sqrt \alpha} \widetilde{W}_{kj}^{\sqrt \alpha}\right)}^{\sqrt {\ell^2 - \alpha}} \widetilde{S}_{ij}^{\ell},  \label{eq:Pi_s3} \\
  \Pi^{\ell}_{m1} = \frac{2}{3} \int_0^{\ell^2} \d\alpha \overline{\overline{\rho}^{\sqrt \alpha}  \widetilde{A}_{ll}^{\sqrt \alpha} \widetilde{S}_{ij}^{\sqrt \alpha} }^{\sqrt {\ell^2 - \alpha}} \widetilde{S}_{ij}^{\ell}, \label{eq:Pi_m1} \\
  \Pi^{\ell}_{m2} = \frac{1}{3}\int_0^{\ell^2} \d\alpha \overline{\overline{\rho}^{\sqrt \alpha}  \widetilde{S}_{ij}^{\sqrt \alpha} \widetilde{S}_{ij}^{\sqrt \alpha} }^{\sqrt {\ell^2 - \alpha}} \widetilde{A}_{ll}^{\ell}, \label{eq:Pi_m2} \\
  \Pi^{\ell}_{m3} = - \frac{1}{3}\int_0^{\ell^2} \d\alpha \overline{\overline{\rho}^{\sqrt \alpha}  \widetilde{W}_{ij}^{\sqrt \alpha} \widetilde{W}_{ji}^{\sqrt \alpha} }^{\sqrt {\ell^2 - \alpha}} \widetilde{A}_{ll}^{\ell}, \label{eq:Pi_m3} \\
  \Pi^{\ell}_d = \frac{1}{9}\int_0^{\ell^2} \d\alpha \overline{\overline{\rho}^{\sqrt \alpha}  \widetilde{A}_{ii}^{\sqrt \alpha} \widetilde{A}_{jj}^{\sqrt \alpha} }^{\sqrt {\ell^2 - \alpha}} \widetilde{A}_{kk}^{\ell} \label{eq:Pi_d}.
\end{gather}
\endgroup
\label{eq:Pi_solution}
\end{subequations}
In Eq.~\eqref{eq:Pi_solution_all}, the inter-scale energy transfer is decomposed into three parts: $\Pi^{\ell}_s$, the energy transfer within the solenoidal mode; $\Pi^\ell_m$, the mixed energy transfer between the solenoidal and the dilatational modes; and $\Pi^\ell_{d}$, the energy transfer within the dilatational mode. More precisely, the solenoidal energy transfer has three contributions: $\Pi^{\ell}_{s1}$, $\Pi^{\ell}_{s2}$ and $\Pi^{\ell}_{s3}$, which represent the energy transfer due to the interaction of strain at scale $\ell$ with strain (see Eq.~\eqref{eq:Pi_s1}), vorticity (see Eq.~\eqref{eq:Pi_s2}), and strain-vorticity correlations (see Eq.~\eqref{eq:Pi_s3}) at a scale smaller than or equal to $\ell$, respectively. These terms correspond to $\Pi^{\ell}_{S}$, $\Pi^{\ell}_{\Omega}$ and $\Pi^{\ell}_{c}$ introduced in the incompressible case by \cite{Johnson2020PRL} and \cite{Yang2023SSAVS}. The mixed energy transfer $\Pi^{\ell}_{m}$ is also composed of three terms: $\Pi^{\ell}_{m1}$, $\Pi^{\ell}_{m2}$ and $ \Pi^{\ell}_{m3}$. The first one, $\Pi^{\ell}_{m1}$, see Eq.~\eqref{eq:Pi_m1}, represents the energy transfer resulting from the interaction between strain at scale $\ell$ and the strain-dilatation correlation at a scale smaller than or equal to $\ell$. The terms $\Pi^{\ell}_{m2}$ (see Eq.~\eqref{eq:Pi_m2}), respectively $\Pi^{\ell}_{m3}$ (see Eq.~\eqref{eq:Pi_m3}) describe the energy transfer resulting from the interaction between dilatation at scale $\ell$ and strain, respectively vorticity. Finally, the dilatational energy transfer $\Pi^{\ell}_{d}$ (see Eq.~\eqref{eq:Pi_d}) arises from the interaction between dilatation at scale $\ell$ and dilatation at a scale smaller than or equal to $\ell$. 

We further decompose the inter-scale energy transfer terms as a sum of a scale-local ($l$) and scale non-local ($nl$) contributions, \ie:
\begin{equation}
  \Pi^{\ell}_{x} = \Pi^{\ell}_{x,l} + \Pi^{\ell}_{x,nl},
\end{equation}
where $x$  denotes one of the subscripts $s1$, $s2$, $s3$, $m1$, $m2$, $m3$, or $d$. The scale-local contributions involve only quantities defined at scale $\ell$, while the scale non-local contributions involve velocity gradients at scales finer than $\ell$. We find that $\Pi^{\ell}_{s3,l}=0$, the other local contributions being expressible in terms of the five independent components of the third-order velocity gradient moment discussed in \cite{yang2022general}:
\begin{subequations}
  \begingroup
  \allowdisplaybreaks
  \begin{gather}
    \Pi^{\ell}_{s1,l} = \overline{\rho}^{\ell}  \widetilde{S}_{ik}^{\ell} \widetilde{S}_{jk}^{\ell} \widetilde{S}_{ij}^{\ell} \ell^2,\\
    \Pi^{\ell}_{s2,l} = - \overline{\rho}^{\ell}  \widetilde{W}_{ik}^{\ell} \widetilde{W}_{kj}^{\ell}\widetilde{S}_{ij}^{\ell} \ell^2, \\
    \Pi^{\ell}_{m1,l} = 2\Pi^{\ell}_{m2,l} = \frac{2}{3}\overline{\rho}^{\ell}  \widetilde{S}_{ij}^{\ell} \widetilde{S}_{ij}^{\ell} \widetilde{A}_{ll}^{\ell} \ell^2, \label{eq:Pim1_Pim2} \\
    \Pi^{\ell}_{m3,l} = - \frac{1}{3} \overline{\rho}^{\ell}  \widetilde{W}_{ij}^{\ell} \widetilde{W}_{ji}^{\ell}  \widetilde{A}_{ll}^{\ell} \ell^2, \\
    \Pi^{\ell}_{d,l} = \frac{1}{9}\overline{\rho}^{\ell}  \widetilde{A}_{ii}^{\ell} \widetilde{A}_{jj}^{\ell}  \widetilde{A}_{kk}^{\ell} \ell^2.
\end{gather}
\endgroup
\label{eq:Pi_solution_local}
\end{subequations}

\section{Numerical methods}
\label{sec:DNS}

To analyze the dilatational contribution of compressible turbulence using the approach developed above, we use DNS of freely decaying compressible homogeneous isotropic turbulence (HIT). To this end, we first solved numerically the compressible Navier-Stokes equations. In this section, we briefly introduce the governing equation, the numerical setup and the computational fluid dynamics solver. We also provide elementary information about the simulated flows.

For the non-dimensional Navier-Stokes equations, we select the reference quantities $\rho_0$ for density, $L_0$ for length, $T_0$ for temperature, $u_0$ for velocity, and $\mu_0$ for viscosity. These five reference quantities are grouped into two dimensionless numbers: the Reynolds number $\textrm{Re}=\rho_0 u_0 L_0/\mu_0$, which quantifies the scale separation of the flow, and the Mach number $\textrm{Ma}=u_0/\sqrt{\gamma R T_0}$ that measures the compressibility of the fluid. $R$ is the specific gas constant and $\gamma=1.4$ is the ratio of specific heats. These two dimensionless numbers are set at the start of each run and remain constant throughout the simulation. The non-dimensional compressible Navier-Stokes equations are expressed as:
\begin{subequations}
\begin{gather}
  \ParDeri{\rho}{t} + \ParDeri{\rho u_i}{x_i} = 0, \label{eq:DNS_1}\\
  \ParDeri{\rho u_i}{t} + \ParDeri{}{x_j}\left(\rho u_i u_j + p\delta_{ij}\right) -  \ParDeri{\sigma_{ij}}{x_j} = 0, \label{eq:DNS_2}\\
  \ParDeri{\Ecal}{t} + \ParDeri{}{x_j}\left[\left(\Ecal+p\right)u_j\right] - \ParDeri{}{x_j}\left(\sigma_{ij}u_i - Q_j\right) = 0, \label{eq:DNS_3}
\end{gather}
\end{subequations}
in which $p=\rho T/(\gamma \textrm{Ma}^2)$ is the pressure, 
$\sigma_{ij}=\mu \cdot [\ParDeriH{u_i}{x_j}+\ParDeriH{u_j}{x_i}-\frac{2}{3}\ParDeriH{u_k}{x_k}\delta_{ij}]$ is the non-dimensional viscous stress tensor with the effect of bulk viscosity neglected \citep{PAN2017role}, $\Ecal=\frac{1}{2}u_i u_i + p/(\gamma-1)$ is the dimensionless total energy and $Q_j= - \mu/[\textrm{Pr}\cdot\textrm{Re}(\gamma-1) \cdot \textrm{Ma}^2] \cdot \ParDeriH{T}{x_j}$ is the dimensionless heat flux with a constant Prandtl number $\textrm{Pr}=0.72$. The viscosity follows the Sutherland's law $\mu=T^{3/2} \cdot (T_0+T_{ref})/ \left[(T \cdot T_0+T_{ref})  \cdot \textrm{Re}\right]$.

Equations \eqref{eq:DNS_1}-\eqref{eq:DNS_3} are solved with a high-order finite difference method. Specifically, the convective terms are calculated by a seventh-order low-dissipative monotonicity-preserving scheme \citep{fang2013optimized} such that shock waves in a compressible flow can be captured and the capabilities of resolving small-scale turbulent structures are preserved. The diffusion terms are discretized by a sixth-order compact central scheme \citep{Lele1994} with a domain decoupling scheme for parallel computation \citep{fang2019improved}. The time integration is computed by a three-step third-order total variation diminishing Runge–Kutta method \citep{gottlieb1998total}. For the flow simulations in this study, we utilize the open-source solver ASTR, which has been extensively validated in DNSs of various compressible turbulent flows with and without shock waves \citep{fang2013optimized,fang2014investigation,fang2015direct,fang2016relation,fang2020turbulence,yang2022general}. 



The computational domain for our numerical simulations is a $(2\pi)^3$ cube with periodic conditions in all three directions, discretized with a $512^3$ mesh. The initial velocity field is divergence-free, with an energy spectrum $E(k)=Ak^4 \exp(-2k^2/k_0^2)$, $k_0=4$ being the wavenumber where the energy spectrum peaks and $A$ determining the initial energy. The density, $\rho$, pressure, $p$, and temperature, $T$, are all initialized as constant, following the `IC4' configuration of \cite{Samtaney2001}. The Reynolds numbers based on the Taylor microscale is defined as $R_\lambda = \aavg{\rho} u' \lambda/ (\sqrt{3}\aavg{\mu})$ and the turbulent Mach number as $Ma_t = \textrm{Ma} \cdot u'/\aavg{\sqrt{\gamma R T}}$. In the descriptions above, $u'=\sqrt{\aavg{u_i u_i}}$ is the root-mean-square velocity and $\lambda = u'/ \aavg{\ParDeriH{u_1}{x_1}^2+\ParDeriH{u_2}{x_2}^2+\ParDeriH{u_3}{x_3}^2}^{1/2}$ is the Taylor microscale. We simulated in total eight flows, with four different initial turbulent Mach numbers: $Ma_t=1.00$, $0.65$, $0.32$, $0.10$ (referred to as 1, 2, 3 and 4 respectively) and 2 different initial Taylor-scale Reynolds numbers $R_\lambda=626.21$, $313.11$ (denoted in the following by H and L respectively). The main parameters characterized the simulations are summarized in Table~\ref{tab:3D_statistics}. Although simulated with a compressible Navier-Stokes solver, the flows 4L and 4H can effectively be considered as incompressible, given that their Mach numbers are extremely small. In the following, we use the initial large-eddy-turnover time $\tau_0 = \left(\int_{0}^{\infty}E(k)/k \d k \right)/u'^3$ to normalize time in the simulations.

\begin{figure}
  \centering
    (\textit{a})\includegraphics[width=0.46 \textwidth]{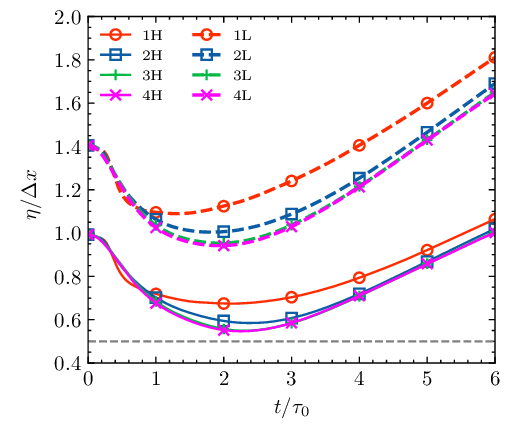}
    (\textit{b})\includegraphics[width=0.46 \textwidth]{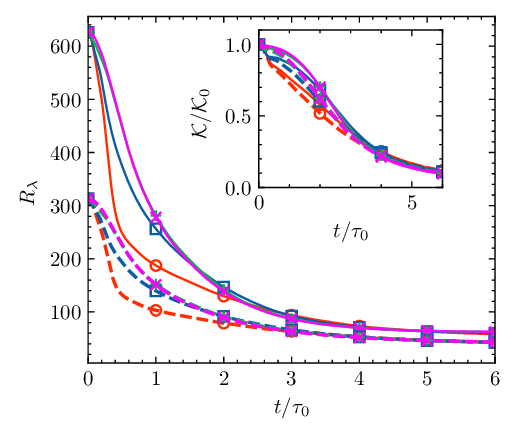}\\
    (\textit{c})\includegraphics[width=0.46 \textwidth]{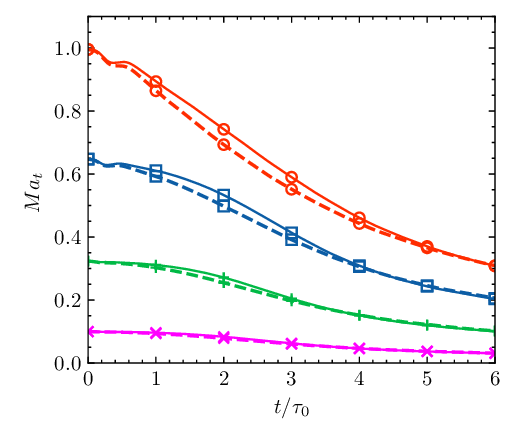}
    (\textit{d})\includegraphics[width=0.46 \textwidth]{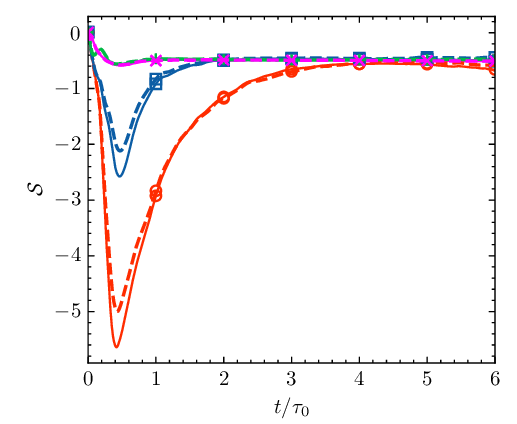}
  \caption{Evolution of global statistics. The solid line and dashed line refer to high and low Reynolds number runs, respectively. Red, blue, green and magenta lines refer to runs 1, 2, 3 and 4, respectively.
  (\textit{a}) The Kolmogorov scale divided by the grid length $\Delta x = 2\pi /512$, the horizontal dashed line marks the threshold $0.5$, (\textit{b}) the Reynolds number on the Taylor microscale $R_\lambda$ and the turbulent kinetic energy normalized by its initial value $\Kcal/\Kcal_0$ (shown in the inset), (\textit{c}) the turbulent Mach number $Ma_t$, (\textit{d}) the velocity gradient skewness $\Scal$.}
\label{fig:3D_evolution}
\end{figure}

\begin{table}
  \begin{center}
\def~{\hphantom{0}}
  \begin{tabular}{lccccc}
    Run  & Initial $R_\lambda$   &   Initial $Ma_t$ & Initial $\eta/\Delta x$ & Minimum $\eta/\Delta x$ & $t_{\eta,\min{}}/\tau_0$ \\[3pt]
       1H  & 626.21 & 1.00 & 0.99 & 0.67 &2.13\\
       2H  & 626.21 & 0.65 & 0.99 & 0.58 &2.38\\
       3H  & 626.21 & 0.32 & 0.99 & 0.55 &2.30 \\
       4H  & 626.21 & 0.10 & 0.99 & 0.55 &2.25 \\
       1L  & 313.11 & 1.00 & 1.40 & 1.09 &1.29 \\
       2L  & 313.11 & 0.65 & 1.40 & 1.00 &1.81\\
       3L  & 313.11 & 0.32 & 1.40 & 0.95 &1.94\\
       4L  & 313.11 & 0.10 & 1.40 & 0.94 &1.93\\
  \end{tabular}
  \caption{Numerical setting and basic statistics}
  \label{tab:3D_statistics}
  \end{center}
\end{table}

Figure \ref{fig:3D_evolution} illustrates the evolution of global statistics. The Kolmogorov scale is defined by $\eta=(\aavg{\mu/\rho}^3/\varepsilon)^{1/4}$, with the energy dissipation rate defined as $\varepsilon = \aavg{\sigma_{ij}\mathring S_{ij}/\rho}$ and $\mathring S_{ij} = (\partial u_i /\partial x_j +\partial u_j / \partial x_i)/2$. As shown in figure \ref{fig:3D_evolution} (\textit{a}), the Kolmogorov scale first decreases due to the development of strong compression. After reaching its minimum at time $t_{\eta,\min{}}$, the Kolmogorov scale $\eta$ subsequently increases as the flow transitions into a decaying regime. After $t_{\eta,\min{}}$, figure \ref{fig:3D_evolution} (\textit{b}-\textit{c}) shows that the turbulent kinetic energy $\Kcal = \aavg{\rho u_i u_i}/2$, the Taylor-scale Reynolds number $R_\lambda$ and the turbulent Mach number $Ma_t$ all decay monotonically. Thus, $t_{\eta,\min{}}$ can be regarded as the starting time of the decay. The initial Kolmogorov scale, the minimum Kolmogorov scale, and the time of the minimal Kolmogrov scale $t_{\eta,\min{}}$ are presented in Table~\ref{tab:3D_statistics}, $\Delta x$ being the grid spacing. For all runs, $\eta/\Delta x$ remains above $0.5$, ensuring adequate resolution.  We also found that, during the decay, the integral scale $L= u'^3/(\varepsilon/\aavg{\rho})$ remains less than one-third of the size of the computational domain (not shown), ensuring that the largest turbulent structures are adequately contained within the simulation box.

Figure~\ref{fig:3D_evolution}(\textit{b}) shows that the turbulent kinetic energy $\Kcal$ and the Taylor-scale Reynolds number $R_\lambda$ all decay monotonically with time. Furthermore, while the curves of $\Kcal$ and $R_\lambda$ deviate from each other during the initial stages, comparing flows at the same initial Reynolds number (\ie, runs 1H, 2H, 3H, 4H, and runs 1L, 2L, 3L, 4L) demonstrates that the curves collapse at later time, during the decay. This observation suggests that, over the range considered here, compressibility does not significantly impact global dissipation during the decay stage. Similarly, figure \ref{fig:3D_evolution}(\textit{c}) shows that the time dependence of $Ma_t$ does not depend much on the initial Reynolds number, but rather solely on the initial Mach number. 
Nevertheless, in runs 1H, 2H, 3H, 1L, 2L, and 3L, despite the relatively small turbulent Mach number towards the end of the simulation, compressible effects persist in the flow field with the presence of spatially distributed shocklets in the flow. In comparison, in our simulations 4H and 4L, the flow remains incompressible throughout the calculation. The skewness of velocity gradient, $\Scal = \aavg{\ParDeriH{u_1}{x_1}^3+\ParDeriH{u_2}{x_2}^3+\ParDeriH{u_3}{x_3}^3}/\Bigl(\aavg{\ParDeriH{u_1}{x_1}^2}^{3/2}+\aavg{\ParDeriH{u_2}{x_2}^2}^{3/2}+\aavg{\ParDeriH{u_3}{x_3}^3}^{3/2}\Bigr)$, shown in Fig.~\ref{fig:3D_evolution}(\textit{d}), exhibits a sharp negative peak for runs 1L, 2L, 1H, and 2H at approximately $t/\tau_0 \approx 0.35$, reflecting the transient from the initial divergence-free velocity field to a compressible flow field. In contrast, for runs 3H, 3L, 4H, and 4L, the skewness decreases during the initial stage. Once the turbulent regime is well established, the skewness remains nearly constant at approximately $- \Scal \approx 0.5$, which aligns with the theoretical estimates for incompressible turbulence as described in~\cite{jian1994skewness}. Overall, our results are consistent with the scaling proposed by~\cite{Donzis2020scaling}.

\section{Numerical study of the energy flux}
\label{sec:num_energy_flux}

In this section, we begin by validating the SGS stress integral expression Eq.~\eqref{eq:tauij_solution}, and we study the total inter-scale energy transfers, see \S \ref{subsubsec:validation}. Furthermore, in \S \ref{subsubsec:contribution}, we present a systematic analysis for the solenoidal and dilatational contributions. We also investigate the scaling of the dilatational contribution as a function of the turbulent Mach number in \S \ref{subsubsec:scaling}. Finally, we apply our results to the LES and discuss the limitations of the current SGS viscosity model in \S \ref{subsubsec:SGSweakness}.

\subsection{Validation of Eq.~\eqref{eq:tauij_solution} and study of total inter-scale energy transfer}
\label{subsubsec:validation}

\begin{table}
  \begin{center}
\def~{\hphantom{0}}
  \begin{tabular}{lcccccc}
       Run & $R_\lambda$ & $Ma_t$ & $\eta/\Delta x$ & $\lambda/\eta$ & $L/\eta$ & $\Scal$ \\[3pt]
       1H & 59.68 &0.33 & 1.00 &14.74 & 181.92 &-0.62 \\
       2H & 61.56 &0.22 & 0.96 &15.12 & 187.21 &-0.50 \\
       3H & 62.87 &0.11 & 0.95 &15.56 & 186.35 &-0.49 \\
       4H & 63.06 &0.03 & 0.94 &15.62 & 185.98 &-0.51 \\
       1L & 43.95 &0.33 & 1.72 &12.67 & 114.31 &-0.56 \\
       2L & 44.24 &0.22 & 1.60 &12.89 & 112.79 &-0.46 \\
       3L & 43.90 &0.11 & 1.56 &13.00 & 108.68 &-0.48 \\
       4L & 44.13 &0.03 & 1.56 &13.05 & 109.44 &-0.51 \\
  \end{tabular}
  \caption{Global statistics at the moment $t/\tau_0 = 5.59$.}
  \label{tab:3D_559_statistics} 
  \end{center}
\end{table}

~~~~In this subsection and in the next one, we focus mostly on the flow field at a single moment in the stable decaying regime, namely $t/\tau_0 = 5.59$, to study the SGS stress and inter-scale energy transfer. Table \ref{tab:3D_559_statistics} lists some of the essential global statistical quantities at the selected time. Note that at this moment, the turbulent Mach numbers $Ma_t$ in all runs satisfy $Ma_t \lesssim 0.3$ and correspond to the low-$Ma_t$ regime \citep{JagannathanDonzis2016scaling}.

\begin{table}
  \begin{center}
  Average relative errors(\%) \\
\def~{\hphantom{0}}
  \begin{tabular}{lccccccc}
      $\ell$ & $\tau^\ell_{11}$ &  $\tau^\ell_{22}$  &  $\tau^\ell_{33}$ & $\tau^\ell_{12}$ & $\tau^\ell_{13}$ & $\tau^\ell_{23}$ & $||\tau^\ell||^2_F = \tau^\ell_{ij}\tau^\ell_{ij}$ \\[3pt]
$4.90 \eta$& 0.43 & 0.43 & 0.43& 0.34 & 0.34& 0.34 & 0.71\\
$15.58 \eta$& 0.86 & 0.87 & 0.88& 0.78 & 0.78& 0.76 & 1.66\\
  \end{tabular}
  \caption{Average relative errors (\%) of the SGS stress calculated with the integral expression $\tau^\ell_{ij,\text{int}}$ with respect to the SGS stress calculated with the definition $\tau^\ell_{ij,\text{def}}$, for the run 1H at $t/\tau_0 = 5.59$.}
  \label{tab:3D_relative_error}
  \end{center}
\end{table}

\begin{figure}
  \centering
    (\textit{a})\includegraphics[width=0.46\textwidth]{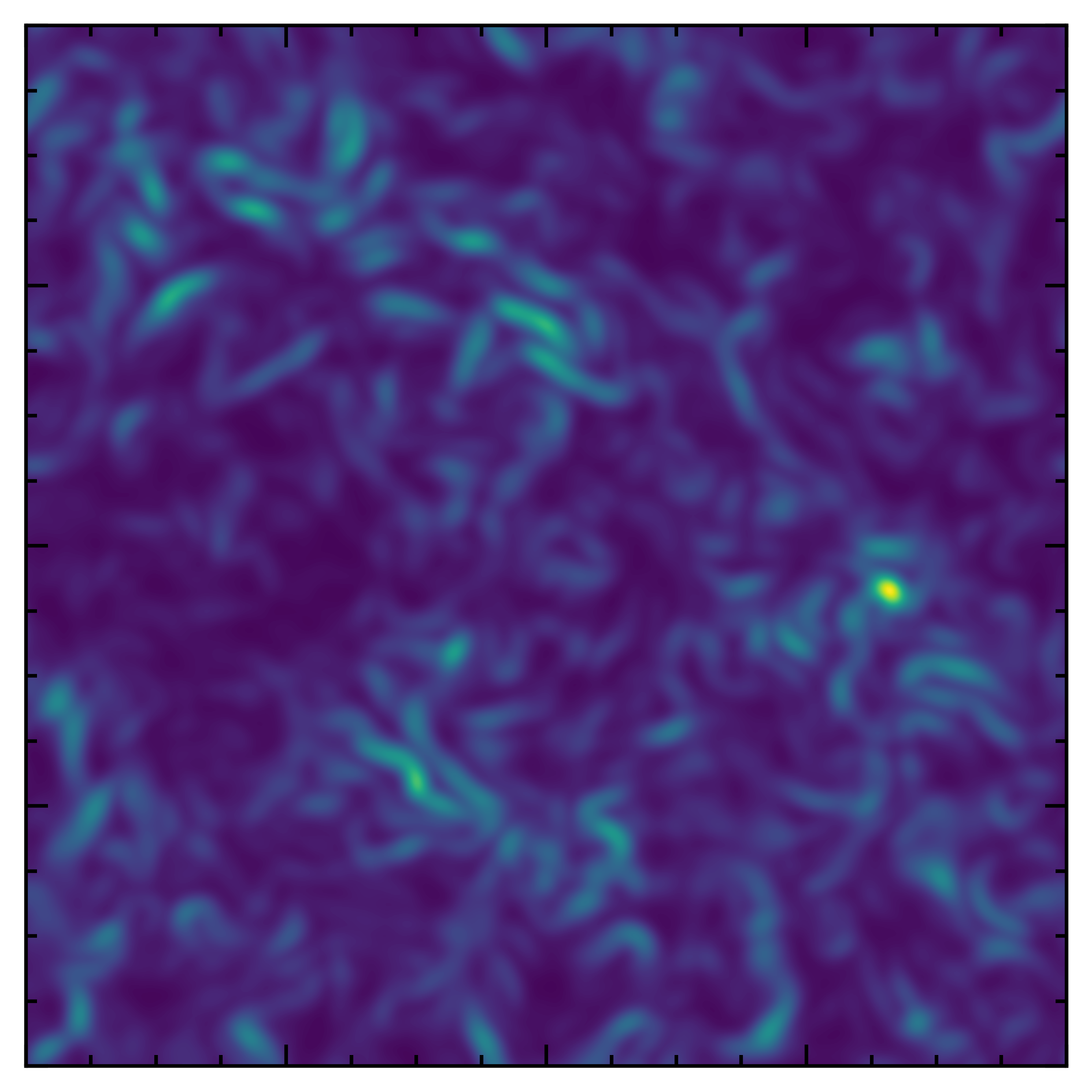}
    (\textit{b})\includegraphics[width=0.46\textwidth]{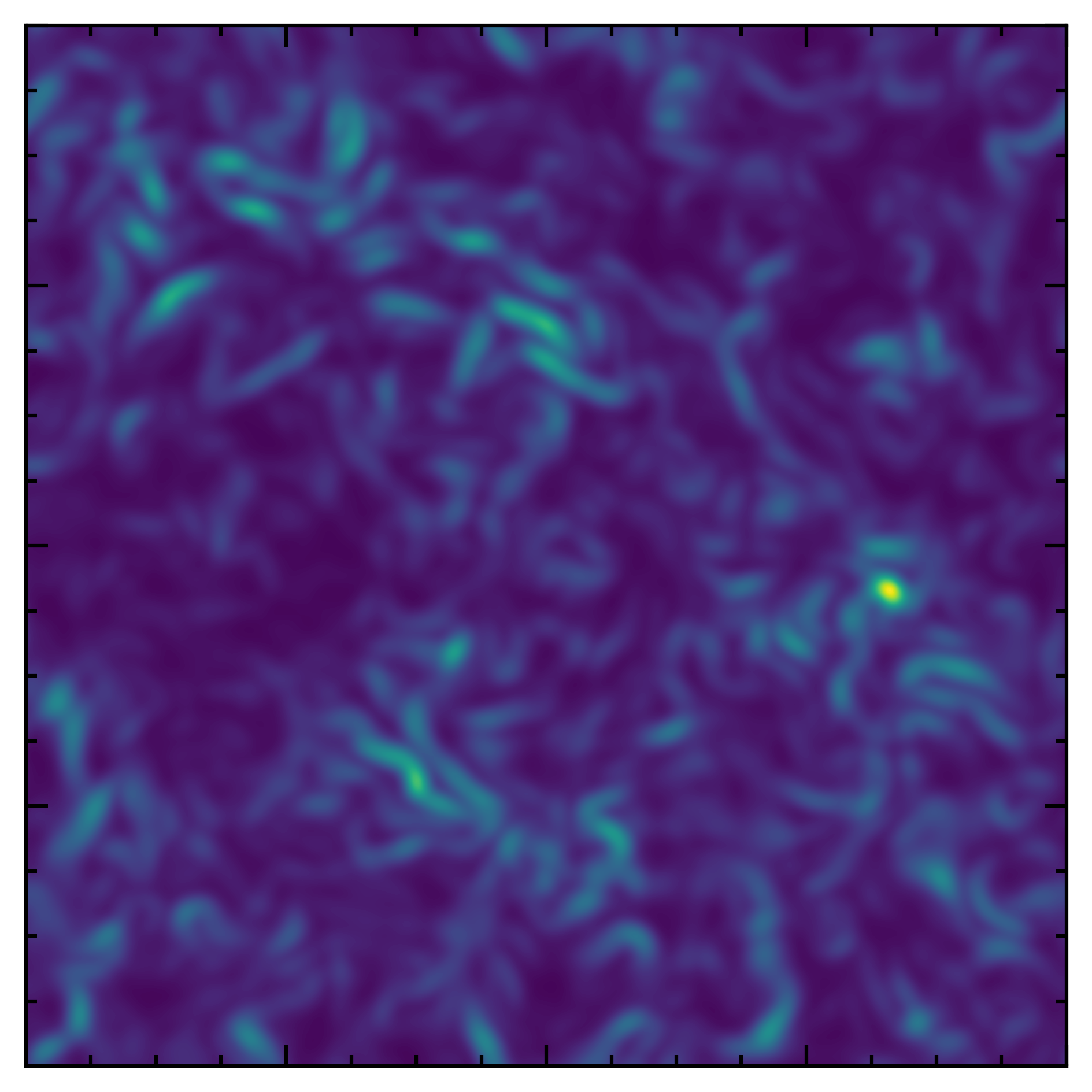}
  \caption{A SGS stress $\tau_{11}^\ell$ of the section $x=\pi$ for the run 1H filtered at the scale $\ell = 4.90 \eta$. Both figures use the same color map. (\textit{a}) With the definition $\tau^\ell_{11,\text{def}}$. (\textit{b}) With the integral expression $\tau^\ell_{11,\text{int}}$.}
\label{fig:3D_SGS_validation}
\end{figure}

First, we validate the integral expression of SGS stress by calculating it using Eqs. \eqref{eq:def_SGS} and \eqref{eq:tauij_solution}. The standard rectangle method is employed for the numerical integration in Eq. \eqref{eq:tauij_solution}, with sufficiently small steps to ensure convergence. The quantities obtained from the integral expressions differ from the definition, Eq.~\eqref{eq:def_SGS} by of the order of $1 \%$. Typical values of the norm of the differences are shown in Table~\ref{tab:3D_relative_error}, at two values of $\ell$. In addition, figure \ref{fig:3D_SGS_validation} shows the SGS stress $\tau_{11}^\ell$ in the plane $x=\pi$ for run 1H at $t/\tau_0 = 5.59$, filtered at the scale $\ell = 4.90 \eta$. In Figure~\ref{fig:3D_SGS_validation}(\textit{a}), we represent the value of $\tau_{11}^\ell$ determined using the definition, Eq.~\eqref{eq:def_SGS}, and in Figure~\ref{fig:3D_SGS_validation}(\textit{b}), the result of the integral expression, Eq.~\eqref{eq:tauij_solution}. The comparison between the two panels demonstrates that the integral expression Eq.~\eqref{eq:tauij_solution} accurately represents the SGS stress, as defined by Eq.~\eqref{eq:def_SGS}. This demonstrates the quality of our numerical results.

\begin{figure}
  \centering
    (\textit{a})\includegraphics[width=0.46 \textwidth]{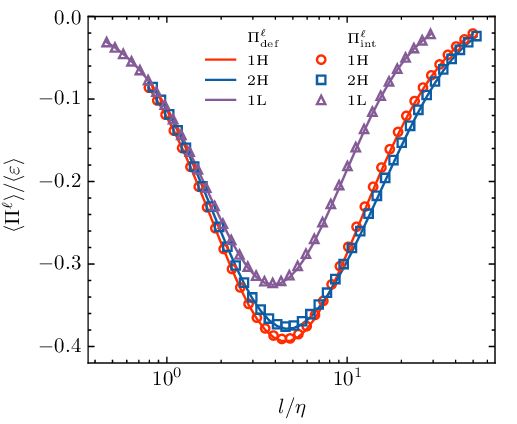}
    (\textit{b})\includegraphics[width=0.46 \textwidth]{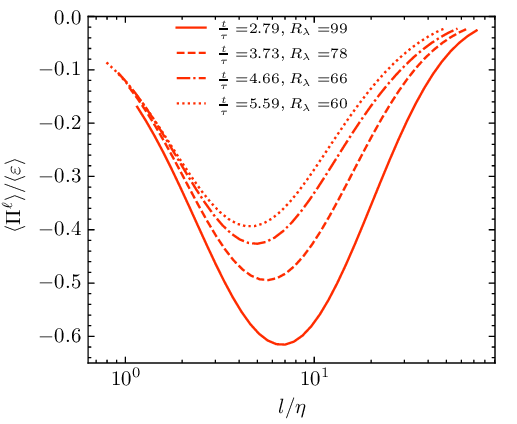} 
  \caption{Total inter-scale energy transfer $\aavg{\Pi^\ell}$ as a function of filtering scale $\ell$. All quantities are normalized by the average energy dissipation rate $\aavg{\epsilon}$. (\textit{a}) At $t/\tau_0 = 5.59$ for runs 1H, 2H and 1L. Solid lines represent the energy transfer calculated from the definition $\Pi^\ell_{\text{def}}$, and symbols represent the energy transfer calculated from the integral expression $\Pi^\ell_{\text{int}}$. (\textit{b}) At different moments for run 1H.}
\label{fig:3D_SGS_Pi}
\end{figure}

Figure \ref{fig:3D_SGS_Pi} shows the dependence of the total inter-scale energy transfer $\aavg{\Pi^\ell}$ on the filtering scale $\ell$ for runs 1H, 2H and 1L, at $Ma_t$ respectively $0.33$, $0.22$ and $0.33$. The solid lines represent the calculations from the definition $\Pi^\ell_{\text{def}} \equiv \tau_{ij}^\ell \frac{\partial \widetilde{u_i}^{\ell}}{\partial x_j}$ and the symbols correspond to the integral expression of Eq.~\eqref{eq:Pi_solution_all}. For all runs and all filtering scales, the solid line and symbols agree with each other to within $2\%$, further validating the numerical integration of the SGS stress. We found a similar agreement for all the other runs (not shown).

Additionally, we observe that the maximum of $|\aavg{\Pi^\ell}/\aavg{\varepsilon}|$ is significantly smaller than 1, whereas in the forced steady-state turbulence simulations of \cite{Johnson2020PRL}~(see Fig.~2 of this article), this ratio is found to be very close to $1$. This can be attributed to the different numerical setups. In \cite{Johnson2020PRL}, energy is continuously injected at large scales, resulting in a forced steady case, with a balance between energy transfer $\aavg{\Pi^\ell}$ and dissipation $\aavg{\varepsilon}$. In contrast, we simulate freely decaying turbulence, where the energy transferred through a certain scale in the inertial range is dissipated later at smaller scales. Consequently, the energy transfer rate at any scale is always lower than the dissipation rate at any given time due to the overall decrease in total energy. Furthermore, viscosity plays a more important role at the low values of $R_\lambda$ considered here, leading to smaller maximum values of $|\aavg{\Pi^\ell}/\aavg{\varepsilon}|$ at lower Reynolds number runs. This is evident when comparing the curves of runs 1H and 1L in \ref{fig:3D_SGS_Pi}(\textit{a}). A similar trend is observed in figure \ref{fig:3D_SGS_Pi}(\textit{b}), which shows the energy transfer at four different time instances for run 1H. As time progresses and the Taylor microscale Reynolds number decreases, the peak value of $\aavg{\Pi^\ell}/\aavg{\varepsilon}$ as a function of $\ell$ also diminishes. We notice in this respect that the values of $R_\lambda \approx 400$ in the work of \cite{Johnson2020PRL} is significantly higher than ours.

\subsection{Solenoidal and dilatational contribution to inter-scale energy flux}
\label{subsubsec:contribution}

~~~~In this subsection, we analyze the solenoidal and dilatational contributions to the energy flux by examining the different terms of the inter-scale energy transfer for high Reynolds number runs over the range of Mach numbers covered by our study: runs 1H ($Ma_t = 0.33$), 2H ($Ma_t = 0.22$), 3H ($Ma_t = 0.11$), and 4H ($Ma_t = 0.03$). Similar trends are observed at lower Reynolds numbers (runs 1L-4L).

\begin{figure}
  \centering
    (\textit{a})\includegraphics[width=0.46 \textwidth]{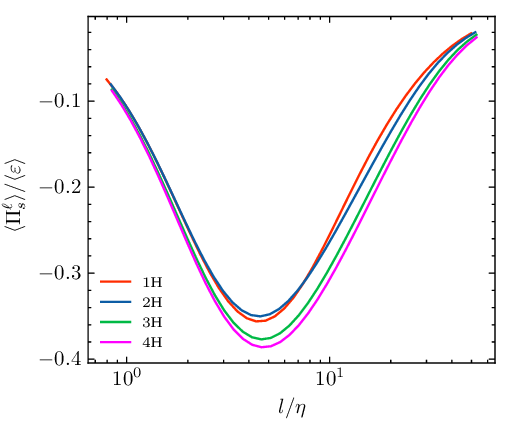}
    (\textit{b})\includegraphics[width=0.46 \textwidth]{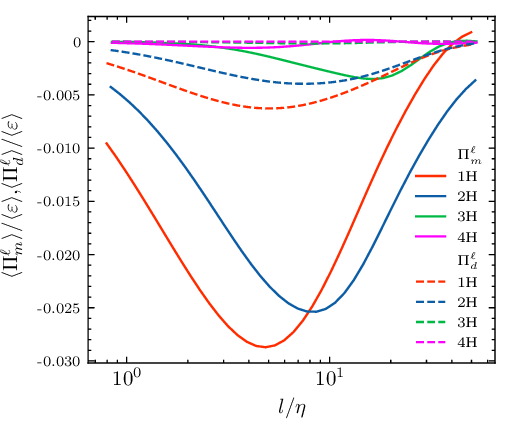}
  \caption{ Average inter-scale energy transfer as a function of filtering scale $\ell$. Red, blue, green and magenta lines refer to runs 1H,2H,3H, and 4H, respectively. All curves are normalized by the average energy dissipation rate $\aavg{\epsilon}$. (\textit{a}) Energy transfer within the solenoidal part $\aavg{\Pi^\ell_{s}}$ at $t/\tau_0 = 5.59$ for the different runs. (\textit{b}) Energy transfer between the solenoidal part and the dilatational part $\aavg{\Pi^\ell_m}$ (solid line) and within the dilatational part $\aavg{\Pi^\ell_{d}}$ (dashed line) at $t/\tau_0 = 5.59$ for the different runs. }
\label{fig:3D_SGS_Pi_term}
\end{figure}

Figure \ref{fig:3D_SGS_Pi_term} shows the dependence on scale, $\ell$, of the various contributions to the inter-scale energy transfer. Specifically, panel~(\textit{a}) shows the contribution due to the solenoidal component, $\Pi^\ell_{s}$, and panel~(\textit{b}) the result of the interaction between the solenoidal and the dilatational components, $\Pi^\ell_m$, as well as the contribution induced by the purely dilatational part, $\Pi^\ell_{d}$. As expected, $\aavg{\Pi^\ell_m}$ and $\aavg{\Pi^\ell_{d}}$ are very close to zero for run 4H ($Ma_t = 0.03$), and increase with the turbulent Mach number. For runs 1H, 2H, and 3H, $\aavg{\Pi^\ell_{s}}$ is significantly larger than $\aavg{\Pi^\ell_m}$ and $\aavg{\Pi^\ell_{d}}$, indicating that the energy transfer is mostly due to the solenoidal motion. This is a consequence of the relatively small values of the corresponding turbulent Mach number ($Ma_t \lesssim 0.3$), which is not expected to be valid beyond the low $Ma_t$ regime. Furthermore, all terms are negative at small scales ($\ell \ll L$), demonstrating that in 3D freely decaying compressible turbulence, all terms contribute to an energy cascade from large to small scales.

\begin{figure}
  \centering
  (\textit{a})\includegraphics[width=0.46 \textwidth]{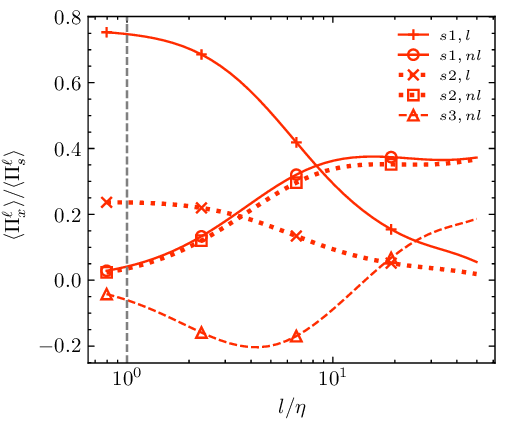}
  (\textit{b})\includegraphics[width=0.46 \textwidth]{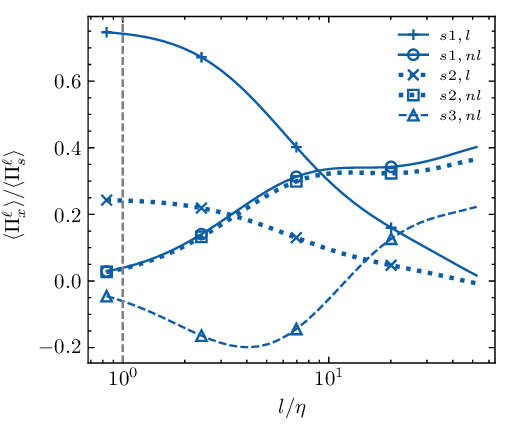}
  (\textit{c})\includegraphics[width=0.46 \textwidth]{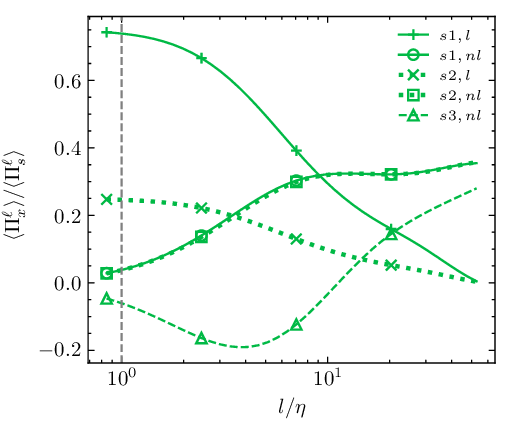}
  (\textit{d})\includegraphics[width=0.46 \textwidth]{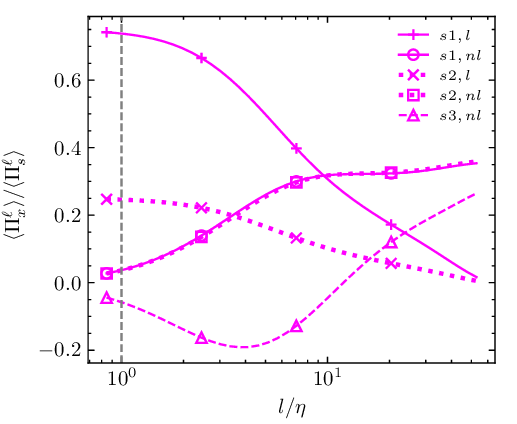}
  \caption{Various terms of the inter-scale energy transfer within the solenoidal mode as a function of filtering scale $\ell$ for the runs at $t/\tau_0 = 5.59$, normalized by the sum of these terms $\aavg{\Pi^\ell_s}$. The gray dashed line indicates the scale $\ell = \eta$. (\textit{a}) Run 1H ($Ma_t = 0.33$), (\textit{b}) Run 2H ($Ma_t = 0.22$), (\textit{c}) Run 3H ($Ma_t = 0.11$), (\textit{d}) Run 4H ($Ma_t = 0.03$). The structure of the energy transfer due to the solenoidal component of the flow is relatively insensitive to the turbulent Mach number over the range covered by our study.}
\label{fig:3D_SGS_S}
\end{figure}

Next, we consider separately all terms contributing to the inter-scale energy transfer, see Section \ref{sec:theory}. Figure \ref{fig:3D_SGS_S} shows the contribution of each term to the inter-scale energy transfer within the solenoidal part, as studied in the incompressible case \citep{Johnson2020PRL,Yang2023SSAVS}. Figure~\ref{fig:3D_SGS_S} exhibits trends similar to those observed in figure 3 of \cite{Johnson2020PRL}. This suggests that in the regime of $Ma_t$ considered here, compressibility has a small influence on the energy transfer associated with the solenoidal component of the velocity. The Betchov identities~\citep{Betchov_1956} imply that $\aavg{\Pi^\ell_{s1,l}} = 3 \aavg{\Pi^\ell_{s2,l}}$, for the scale-local terms in homogeneous flows. For scale non-local interactions, the approximation of $\aavg{\Pi^\ell_{s1,nl}} \approx \aavg{\Pi^\ell_{s2,nl}}$ holds for all $\ell$, consistent with the numerical observations of \cite{Johnson2020PRL}, see also the theoretical discussion of \cite{Yang2023SSAVS}. Furthermore, when $\ell \lesssim \eta$, the energy transfer is dominated by local terms,  the non-local terms being comparatively small. Specifically, at $\ell = \eta$, the three non-local terms $\Pi^\ell_{s1,nl},\Pi^\ell_{s2,nl},$ and $\Pi^\ell_{s3,nl}$ account for approximately $4\%, 4\%,$ and $-6\%$ of the total solenoidal energy transfer, respectively. This indicates that non-local energy transfer at scales below $\eta$ is very weak. Consequently, numerical simulations with a grid size of the order of $\eta$ correctly represent the energy transfer due to the solenoidal energy. This is consistent with the classical spatial resolution criterion for DNS, \ie,   $\Delta x \lesssim 2 \eta$.

However, at larger scales ($\ell \gg \eta$), non-local contributions dominate the local ones, accounting for almost the total solenoidal inter-scale energy transfer. This result differs from \cite{Johnson2020PRL}, which found that $\aavg{\Pi^\ell_{s,l}} \approx \aavg{\Pi^\ell_{s,nl}}$. This discrepancy may be attributed to the different Reynolds numbers (with $R_\lambda < 100$ in our work and $R_\lambda = 400$ in~\cite{Johnson2020PRL}) and the different flow types (freely decaying turbulence here versus forced steady flow in \cite{Johnson2020PRL}), which may influence the structure of the large-scale flow \citep{yang2018generalized}.

\begin{figure}
  \centering
  (\textit{a})\includegraphics[width=0.46 \textwidth]{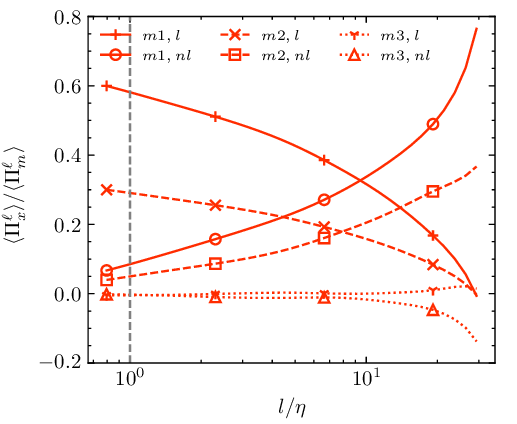}
  (\textit{b})\includegraphics[width=0.46 \textwidth]{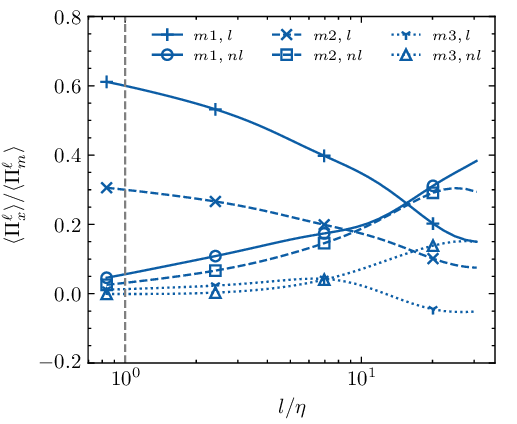}
  \caption{Terms of the inter-scale energy transfer between the solenoidal part and the dilatational part as a function of filtering scale $\ell$ at $t/\tau_0 = 5.59$, normalized by the sum of these terms $\aavg{\Pi^\ell_m}$. The gray dashed line shows the scale $\ell = \eta$.(\textit{a}) Run 1H ($Ma_t = 0.33$), (\textit{b}) run 2H ($Ma_t = 0.22$).}
\label{fig:3D_SGS_SD}
\end{figure}

Figure \ref{fig:3D_SGS_SD} shows the relative contributions of each term to the inter-scale energy transfer due to the interaction between the solenoidal and dilatational components. The structure of the solution is shown for run 1H, panel (\textit{a}) and for run 2H, panel (\textit{b}). For $\ell \lesssim \eta$, the local terms $\aavg{\Pi^\ell_{m1,l}}$ and $\aavg{\Pi^\ell_{m2,l}}$ dominate the energy transfer, with a ratio between the local contributions, defined by Eq.~\eqref{eq:Pim1_Pim2}, $\aavg{\Pi^\ell_{m1,l}}:\aavg{\Pi^\ell_{m2,l}} = 2:1$. Notably, at $\ell = \eta$, the contributions of the non-local terms, $\Pi^\ell_{m1,nl}$ and $\Pi^\ell_{m2,nl}$, are approximately $8\%$ and $5\%$ of the total solenoidal-dilatational energy transfer, respectively. This is larger than the non-local contributions induced by the solenoidal component. At large scales ($\ell \gg \eta$), the inter-scale energy transfer is dominated by the non-local term $\aavg{\Pi^\ell_{m1,nl}}$ involving strain and dilatation. The interaction between dilatation and vorticity, described by the term $\aavg{\Pi^{\ell}_{m3}}$, contributes very little to the energy transfer. This is consistent with previous observations that vorticity and dilatation are nearly uncorrelated in compressible homogeneous turbulence and have a small contribution to energy transfer~\citep{erlebacher1993pof, Wang2012JFM, yang2022general}. The relatively large values of $\aavg{\Pi^\ell_{m3,l}}$ and $\aavg{\Pi^\ell_{m3,nl}}$ in figure \ref{fig:3D_SGS_SD}(\textit{b}) at large scales may be a special property of the individual configuration considered here.

\begin{figure}
  \centering
  \includegraphics[width=0.46 \textwidth]{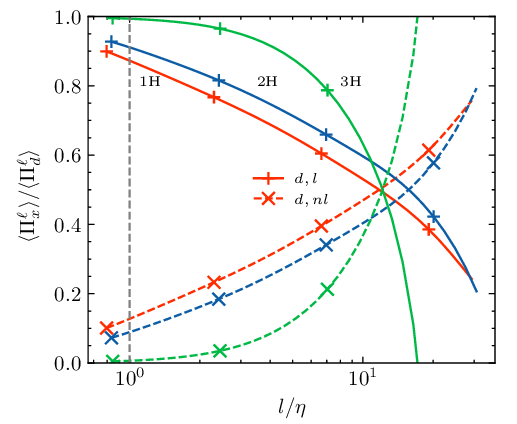}
  \caption{Terms of the inter-scale energy transfer within the dilatational part as a function of filtering scale $\ell$  at $t/\tau_0 = 5.59$, normalized by the sum of these terms $\aavg{\Pi^\ell_d}$. Red, blue and green lines refer to runs 1H,2H, and 3H, respectively. The gray dashed line shows the scale $\ell = \eta$.}
\label{fig:3D_SGS_D}
\end{figure}

Figure \ref{fig:3D_SGS_D} shows the relative contributions of the local and non-local terms to the inter-scale energy transfer, $\aavg{\Pi_{d,l}^{\ell}}$ and $\aavg{\Pi_{d,nl}^{\ell}}$, due to the dilatational component. As it was the case for $\aavg{\Pi^\ell_s}$ and $\aavg{\Pi^\ell_m}$, the local contribution dominates at small scales, for $(\ell \lesssim \eta)$, and decreases when $\ell$ increases, while the non-local contribution is close to $0$ at small scales and increases with $\ell$. Notably, at $\ell = \eta$, $\aavg{\Pi^\ell_{d,nl}}/\aavg{\Pi^\ell_{d}}$ is $13\%$ for run 1H, $9\%$ for run 2H and less than $1\%$ for the run 3H. This ratio manifestly increases with the Mach number, $Ma_t$, and is greater than $\left(\aavg{\Pi^\ell_{s1,nl}}+\aavg{\Pi^\ell_{s2,nl}}+\aavg{\Pi^\ell_{s3,nl}}\right)/\aavg{\Pi^\ell_{s}}$, which is the non-local contribution observed in the solenoidal energy transfer. This signifies that the non-local motions at scales smaller than $\eta$ play a more significant role in the dilatational energy transfer at $\eta$. The implication for numerical simulations is that even with a grid size of the order of $\eta$, uncertainties in $\aavg{\Pi^\eta_{d,nl}}$ may significantly affect the energy transfer due to the dilatational component of the velocity. This implies, in turn, that the classical spatial resolution criteria based on the Kolmogorov scale may not be sufficient to accurately capture the effect of the dilatational terms.

\subsection{Scaling of dilatational contribution with Mach number}
\label{subsubsec:scaling}

~~~~In this subsection, we study the relationship between the dilatational component of the inter-scale energy transfer and the turbulent Mach number.

\begin{figure}
  \centering
  (\textit{a})\includegraphics[width=0.46 \textwidth]{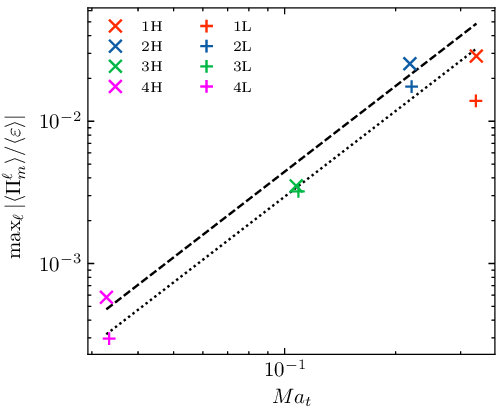}
  (\textit{b})\includegraphics[width=0.46 \textwidth]{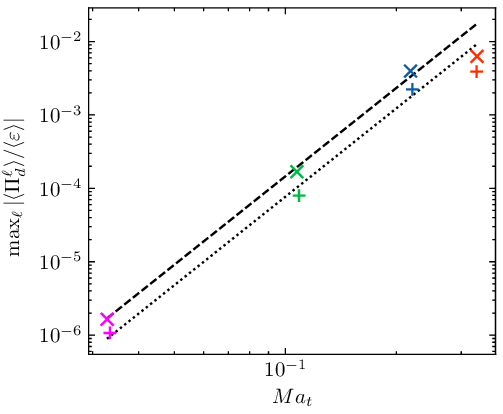} \\

  \caption{The maximum inter-scale energy transfer $\max_\ell |\aavg{\Pi^{\ell}_x}/\aavg{\varepsilon}|$ as a function of turbulent Mach number $Ma_t$ at $t/\tau=5.59$. Both axis are in logarithmic scale. `x' symbol and `+' symbol refer to high and low Reynolds number runs, respectively. Red, blue, green and magenta symbols refer to runs 1, 2, 3, and 4, respectively. The black dashed line and the black dotted line are the linear regression of data of runs 2H, 3H, 4H and 2L, 3L, 4L, respectively. (\textit{a}) Inter-scale energy transfer caused by the solenoidal-dilatational interaction $\max_\ell |\aavg{\Pi^{\ell}_m}/\aavg{\varepsilon}|$.  (\textit{b}) Inter-scale energy transfer caused by the dilatational motion $\max_\ell |\aavg{\Pi^{\ell}_d}/\aavg{\varepsilon}|$. }
\label{fig:3D_scaling}
\end{figure}

For each run, we compute the maximum inter-scale energy transfer $|\aavg{\Pi^{\ell}_x}/\aavg{\varepsilon}|$ with $x=m$ and $d$ as a function of $\ell$. Figures \ref{fig:3D_scaling}(\textit{a}) and \ref{fig:3D_scaling}(\textit{b}) show this maximum value at $t/\tau=5.59$ as a function of the turbulent Mach number. The data in figure \ref{fig:3D_scaling}(\textit{a}) can be fitted as $\max_\ell |\aavg{\Pi^{\ell}_m}/\aavg{\varepsilon}| \approx 0.44 Ma_t^2$ for runs 2H, 3H and 4H, and $\max_\ell |\aavg{\Pi^{\ell}_m}/\aavg{\varepsilon}| \approx 0.30 Ma_t^2$ for runs 2L, 3L and 4L. Similarly, the data in figure \ref{fig:3D_scaling}(\textit{b}) can be fitted as $\max_\ell |\aavg{\Pi^{\ell}_d}/\aavg{\varepsilon}| \approx 1.46 Ma_t^4$ for runs 2H, 3H and 4H, and as 
$\max_\ell |\aavg{\Pi^{\ell}_d}/\aavg{\varepsilon}| \approx 0.77 Ma_t^4$ for runs 2L, 3L and 4L. Both $\max_\ell |\aavg{\Pi^{\ell}_m}/\aavg{\varepsilon}|$ and $\max_\ell |\aavg{\Pi^{\ell}_d}/\aavg{\varepsilon}|$ are increasing functions of the turbulent Mach number and vary respectively as $\sim Ma_t^2$ and $\sim Ma_t^4$. The prefactors of the scaling laws appear to be increasing with the Reynolds numbers. Notably, runs 1H and 1L deviate from the regression lines, suggesting that the scaling laws apply only in the  low-$Ma_t$ regime of compressible turbulence, with a turbulent Mach number $Ma_t \lesssim 0.3$.

\subsection{Examination on the eddy-viscosity assumption in LES modeling}
\label{subsubsec:SGSweakness}
~~~~When the turbulent Mach number is relatively low ($Ma_t \lesssim 0.3$),  the solenoidal modes are expected to contain most of the turbulent kinetic energy~\citep{Sagaut2013LES}. In this low-$Ma_t$ regime, the Boussinesq eddy-viscosity assumption~\citep{boussinesq1877essai} could be extended to the compressible LES, suggesting the following expression for the SGS stress~\citep{Sagaut2013LES}:
\begin{equation}
  \tau_{ij}^\ell=-\bar{\rho}^\ell \nu_{sgs} \left(\ParDeri{\tilde{u}_i^{\ell}}{x_j}+\ParDeri{\tilde{u}_j^{\ell}}{x_i}\right) = -2\bar{\rho}^{\ell} \nu_{sgs} \left(\widetilde{S}_{ij}^{\ell} +\frac{\widetilde{A}_{ll}^{\ell} \delta_{ij}}{3}  \right),
\end{equation}
used in many classical LES models, such as the Smagorinsky model~\citep{smagorinsky1963general}, the mixed scales model~\citep{yoshizawa1986statistical}, and the structure function model~\citep{sagaut1995simulations}. 

Here, we discuss a limitation of this assumption in LES of compressible flows. By substituting the eddy-viscosity assumption into the definition of inter-scale energy transfer, we could get the modeled inter-scale energy transfer as:
\begin{equation}
  \Pi^\ell_{\textrm{ev}} =   \underbrace{-2\bar{\rho}^\ell \nu_{sgs}\widetilde{S}_{ij}^{\ell} \widetilde{S}_{ij}^{\ell}}_{\Pi^{\ell}_{s,\textrm{ev}}} \underbrace{-\frac{2}{3} \bar{\rho}^\ell \nu_{sgs}\widetilde{A}_{ll}^{\ell}\widetilde{A}_{ll}^{\ell}}_{\Pi^{\ell}_{d,\textrm{ev}}}, \label{eq:Pi_model_all}
\end{equation}
where the ``ev'' in the subscript signifies eddy-viscosity. The eddy-viscosity assumption is theoretically valid only if the real inter-scale energy transfer and the modeled one are equal. By comparing Eq.~\eqref{eq:Pi_solution_all} and Eq.~\eqref{eq:Pi_model_all}, it can be concluded that there are two requirements for the eddy-viscosity assumption:
\begin{itemize}
  \item For the chosen filtering scale $\ell$, $\aavg{\Pi^\ell_m}$ should be negligible comparing with $\aavg{\Pi^{\ell}}$.
  \item For the chosen filtering scale $\ell$, the ratio of energy transfer $\aavg{\Pi^\ell_d}/\aavg{\Pi^\ell_s}$ and of filtered velocity gradient moment $\aavg{\widetilde{A}_{ll}^{\ell}\widetilde{A}_{ll}^{\ell}}/\aavg{3 \widetilde{S}_{ij}^{\ell} \widetilde{S}_{ij}^{\ell}}$ should be the same.
\end{itemize}

\begin{figure}
  \centering
  (\textit{a})\includegraphics[width=0.46 \textwidth]{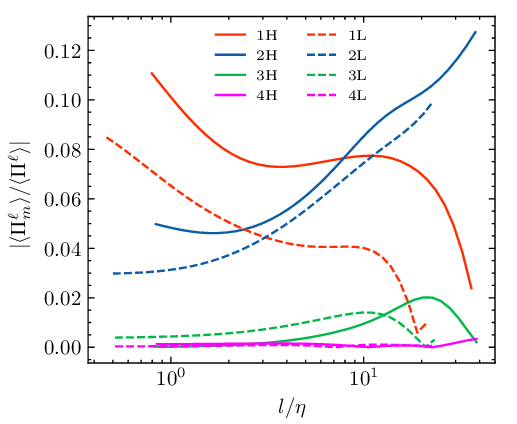}
  (\textit{b})\includegraphics[width=0.46 \textwidth]{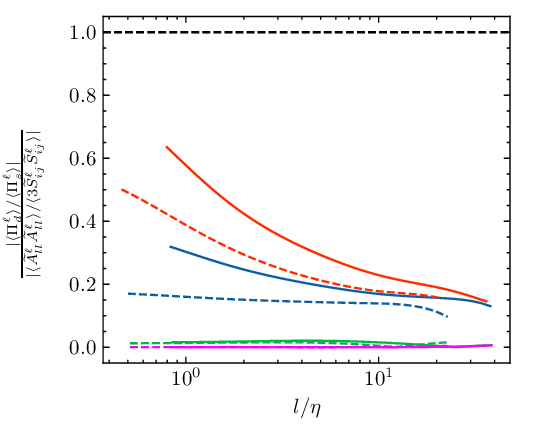}
  \caption{Critical ratios as a function of the filtering scale $\ell$ at $t/\tau_0 = 5.59$. The solid line and dashed line refer to high and low Reynolds number runs, respectively. Red, blue, green and magenta lines refer to runs 1, 2, 3, and 4, respectively. (\textit{a})The ratio of $\aavg{\Pi^\ell_m}/\aavg{\Pi^\ell}$.  (\textit{b}) The ratio of $\left|\aavg{\Pi^\ell_{d}}/\aavg{\Pi^\ell_{s}}\right|/\left|\aavg{\widetilde{A}_{ll}^{\ell}\widetilde{A}_{ll}^{\ell}}/\aavg{3 \widetilde{S}_{ij}^{\ell} \widetilde{S}_{ij}^{\ell}}\right|$. The black dashed line shows the reference value 1.}
\label{fig:3D_SGS_LES}
\end{figure}

For our numerical cases, Figure~\ref{fig:3D_SGS_LES}(\textit{a}) shows the dependence on the filtering scale, $\ell$, of the ratio $\aavg{\Pi^\ell_m}/\aavg{\Pi^\ell}$ obtained from our simulations. For runs 1H, 1L, 2H, and 2L, corresponding to $Ma_t \gtrsim 0.2$, $\aavg{\Pi^\ell_m}$ accounts for more than $4\%$ for the total inter-scale energy transfer, which is not captured by the SGS viscosity model. Besides, figure \ref{fig:3D_SGS_LES}(\textit{b}) shows a discrepancy between the fraction of the modeled energy transfer $\left(\aavg{\Pi^\ell_{d,\textrm{ev}}}/\aavg{\Pi^\ell_{s,\textrm{ev}}}\right)$ and the actual energy transfer $\left(\aavg{\Pi^\ell_{d}}/\aavg{\Pi^\ell_{s}}\right)$. As a result, the eddy-viscosity assumption leads to an overestimation on the dilatational effects. These observations highlight the limitations of the direct application of the eddy-viscosity assumption to compressible flows, and suggest potential directions for improving future SGS modeling techniques. 

\section{Concluding Remarks}
\label{sec:conclution}
In summary, we derived an integral expression for subgrid-scale (SGS) stress of compressible turbulence and introduced a methodology to analyze inter-scale energy transfer. This allows us to formally decompose the kinetic energy transfer into components associated with solenoidal motion, solenoidal-dilatational interaction, and dilatational motion, which are further categorized into scale-local and non-local contributions.

Our theoretical decomposition is validated through DNS of freely decaying compressible turbulence. The results reveal that at moderate turbulent Mach number, $Ma_t \lesssim 0.3$, the energy transfer due to the solenoidal component of the flow follows a pattern similar to that observed in incompressible turbulence, and is not particularly sensitive to the turbulent Mach number, $Ma_t$. Our analysis of the contributions of the different terms for the energy transfer due to the interaction between solenoidal and dilatational, and to the dilatational components shows that non-local dilatational motions smaller than the Kolmogorov scale contribute substantially to the energy transfer. This suggests that traditional spatial resolution criteria of DNS for incompressible turbulence may be insufficient for compressible turbulence, even at moderate turbulent Mach number. Additionally, for weakly compressible turbulence, we observed that $\max_\ell |\aavg{\Pi^{\ell}_m}/\aavg{\varepsilon}|$ and $\max_\ell |\aavg{\Pi^{\ell}_d}/\aavg{\varepsilon}| $ scale with the turbulent Mach number as $Ma_t^2$ and $ Ma_t^4$, highlighting the impact of compressibility on energy transfer dynamics. 
Our study also identifies limitations of the eddy-viscosity assumption when applied to large-eddy simulations (LES) of compressible turbulence. Specifically, this assumption tends to ignore energy transfer associated with solenoidal-dilatational interactions while overestimating dilatational effects. This discrepancy suggests the potential directions for improving future SGS modeling techniques to more accurately capture the complex interactions in compressible turbulence.

\section*{Acknowledgements}
PFY, HX and LF acknowledge support from the National Natural Science Foundation of China (NSFC) under grants 12202452, 11672157, 91852104 and 12372214.
JF acknowledges the UK Engineering and Physical Sciences Research Council (EPSRC) through the Computational Science Centre for Research Communities (CoSeC), and the UK Turbulence Consortium (No. EP/R029326/1).
AP was supported by the French Agence Nationale de la Recherche under Contract No. ANR-20-CE30-0035 (project TILT), and is thankful to the Max Planck Institute for Dynamics and Self-organisation (G\"ottingen, Germany) for its continuous support.

\noindent
\textbf{Declaration of Interest:}
The authors report no conflict of interest.

\bibliographystyle{jfm}
\bibliography{BIB}

\end{document}